\documentclass{article}

\usepackage{PRIMEarxiv}

\usepackage[utf8]{inputenc} 
\usepackage[T1]{fontenc}    
\usepackage{hyperref}       
\usepackage{url}            
\usepackage{booktabs}       
\usepackage{amsfonts}       
\usepackage{nicefrac}       
\usepackage{microtype}      
\usepackage{lipsum}
\usepackage{fancyhdr}       
\usepackage{graphicx}       
\graphicspath{{media/}}     
\usepackage{amsmath}
\title{Hydrological Constraints on Temperature Sensitivities of Precipitation Frequency and Intensity

}

\author{
  Jun Yin\\
  Department of Hydrometeorology \\
  Nanjing University of Information Science and Technology \\
  Nanjing, 210044, China\\
  \texttt{jy12@princeton.edu} \\
   \And
  Amilcare Porporato \\
  Department of Civil and Environmental Engineering \\
  Princeton University \\
  Princeton, 08540, USA\\
  \texttt{aporpora@princeton.edu} \\
}

\begin{document}
\maketitle

\begin{abstract}
To improve understanding of how precipitation frequency and intensity change with
warming, we combine the coupled land--atmosphere water balance with a stochastic
hydrological model that retains explicit dependence on precipitation
frequency and event depth. We find a linear relationship between co-variations of rainfall frequency and intensity, where the slope is a hydrological sensitivity determined by the dryness and storage indices and the intercept is a forcing term associated with changes in moisture convergence, potential evapotranspiration, and effective water-storage capacity. While a given relative change in either rainfall frequency or intensity produces an identical relative change in the mean rate, rainfall intensity additionally influences the runoff coefficient by altering terrestrial water storage relative to rainfall depth. This means frequency and intensity have different impacts on the water cycle, allowing the frequency--intensity slope to depart from -1. Analysis of long-term MOPEX records shows that the forcing term is dominated by moisture convergence and that the frequency--intensity relation becomes steeper from wet to dry regimes,
consistent with the slope predicted by the theory. The framework
provides a physical interpretation of the observed negative covariation
between precipitation frequency and intensity and clarifies the roles
of hydrological partitioning and atmospheric moisture supply in their
responses to warming.
\end{abstract}


\section{Introduction}

Precipitation is a fundamental driver of the terrestrial water cycle,
shaping ecosystems, agriculture, water resources, and natural hazards
\cite{Trenberth2011,Ehtasham2024}. While much attention has been devoted
to changes in mean precipitation and extreme events under climate change
\cite{ipcc_2023}, rainfall intermittency---how often it rains and how
much precipitation falls during wet periods---is also critical for
hydrological and ecological responses
\cite{Knapp2002,Liu2020,Kotz2022,Feldman2024}. For instance, shifts
toward fewer but heavier rainfall events can increase drought stress
between storms while enhancing runoff and flood risk, whereas more
frequent but lighter rainfall may favor soil-water retention and
evapotranspiration \cite{ecohydrology2022}. Understanding how
precipitation frequency and intensity change together is therefore
important for predicting the hydrological consequences of climate
change.

Much of the existing literature has considered the responses of rainfall
frequency and intensity separately. A large body of work has focused on
precipitation extremes, showing that extreme rainfall generally
intensifies with warming at rates broadly related to
Clausius--Clapeyron scaling
\cite{Trenberth1999,Westra2014,OGorman2015}. The frequency of extreme
precipitation events may also increase
\cite{Myhre2019Frequency,Thackeray2022Constraining}. At the same time,
observational and modeling studies show that mean wet-day frequency may
decrease in some regions even as rainfall intensity increases
\cite{Rajah2014,Pendergrass2017,Giorgi2019}. These contrasting responses
point to the need for a framework that treats frequency and intensity
jointly and relates their covariation to the physical processes governing
the coupled water cycle.

\begin{figure}
  \centering
  \includegraphics[width=0.5\linewidth]{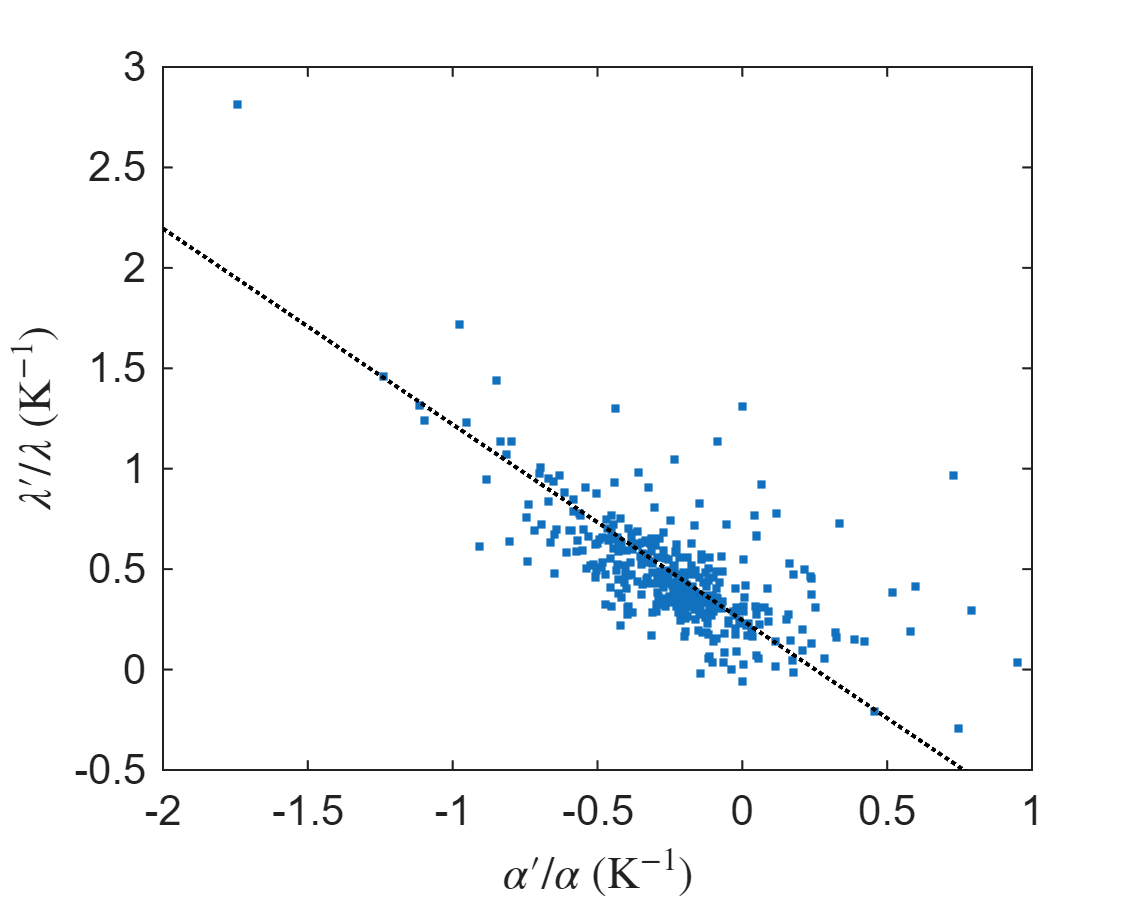}
  \caption{
  Temperature sensitivities of precipitation frequency ($\lambda$) and
  intensity ($\alpha$) estimated from the Model Parameter Estimation
  Experiment (MOPEX) dataset. Each point represents one basin and shows
  the fractional sensitivities $\lambda'/\lambda$ and $\alpha'/\alpha$,
  estimated from 15-year moving-window precipitation statistics regressed
  against the corresponding global-mean temperature over 1948--2003.
  The dashed line denotes the first principal component of the scatter.
  Details of the observational analysis are given in Sections~4 and~5.
  }
  \label{fig:anti}
\end{figure}

Recent work by the authors \cite{yin2026scaling}, based on extensive
rain-gauge observations, identified a robust negative covariation between
changes in rainfall frequency and intensity: regions in which rainfall
becomes less frequent tend, on average, to exhibit increasing rainfall
intensity, and vice versa. A similar pattern is evident in the MOPEX
records considered here (Fig.~\ref{fig:anti}). That work established the
empirical regularity and suggested an important role for hydrological
partitioning. What remains unclear is why this frequency--intensity
trade-off arises, what determines its slope, and how changes in moisture
supply modify the relationship.

Here we ask whether the observed covariation between changes in precipitation frequency and intensity can be understood within the framework of the coupled land--atmosphere water balance, and specifically whether its structure can be decomposed into distinct hydrological and atmospheric contributions. This decomposition is non-trivial: in conventional statistical approaches, land-to-atmosphere feedbacks are intrinsically weaker and noisier than the dominant atmospheric forcing imposed on the land surface. As a result, causal relationships are difficult to disentangle from mere co-variability and land-surface signals are frequently masked by atmospheric variability \cite{Koster2004,Tuttle2016}. To overcome this limitation, here we combine the long-term water balance with a stochastic hydrological model that preserves rainfall frequency and event depth as separate quantities, enabling us to theoretically explore how each independently influences the partitioning of precipitation between evapotranspiration and hydrological export.

The analysis therefore focuses on the physical origin of the
frequency--intensity trade-off, rather than treating it only as an
empirical correlation. Particular attention is given to the role of
finite water storage, which makes changes in rainfall event depth
hydrologically distinct from changes in rainfall frequency even when
their effects on mean precipitation are similar.

This study aims to address three related questions. First, what
hydrological properties determine the slope of the frequency--intensity
relationship? Second, how do changes in atmospheric moisture supply and
other forcing terms shift that relationship? Third, does the resulting
theoretical structure help explain the different temperature responses
observed across wet and dry hydrological regimes?

The paper proceeds from the physical formulation to its observational
evaluation. Section~2 introduces the coupled water balance and stochastic
hydrological model, while Section~3 develops the resulting
frequency--intensity constraint and summarizes its behavior across
hydrological regimes. Sections~4 and~5 examine the forcing terms and
temperature sensitivities using hydrometeorological records. Section~6
summarizes the main conclusions, with detailed derivations and asymptotic
results provided in the Appendices.

\section{Coupled Water Balance and Stochastic Hydrological Partitioning}
\label{sec:wbframe}

We first introduce the coupled land--atmosphere water balance and the
stochastic hydrological model used to describe long-term water-budget
partitioning. Together, they provide the basis for the analytical
development that follows.

\subsection{Long-Term Water Balance}

\begin{figure}
    \centering
    \includegraphics[width=0.3\linewidth]{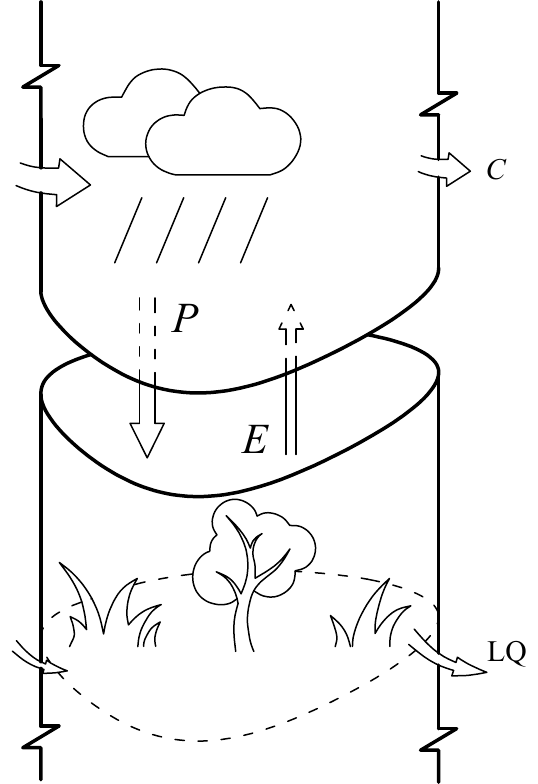}
    \caption{Schematic of the coupled atmospheric and terrestrial water balances over a common horizontal domain. Precipitation $P$ transfers water from the atmosphere to the land surface, evapotranspiration $E$ returns water to the atmosphere, $C$ denotes net atmospheric moisture convergence, and $\mathrm{LQ}$ represents terrestrial water export through runoff and deep drainage. Redrawn from \cite{Yin2019looking}.}
    \label{fig:domain}
\end{figure}

Consider an atmospheric column and the underlying land surface over the
same horizontal domain (Fig.~\ref{fig:domain}). The vertically integrated
atmospheric water balance is
\begin{equation}
\label{eq:dwdt}
    \frac{dW}{dt}=-P+E+C,
\end{equation}
where $W$ is atmospheric water storage, $P$ is precipitation, $E$ is
evapotranspiration, and $C$ is net atmospheric moisture convergence,
taken as positive for convergence into the domain.

The corresponding terrestrial water balance is
\begin{equation}
\label{eq:dsdt}
    w_0\frac{ds}{dt}=P-E-\mathrm{LQ},
\end{equation}
where $s$ is relative soil moisture, $w_0$ is the effective water-storage
capacity, and $\mathrm{LQ}$ denotes water leaving the active terrestrial
reservoir through runoff and deep drainage.

For the long-term climatic balances considered here, changes in
atmospheric and terrestrial water storage are assumed negligible relative
to the mean fluxes. Equations~(\ref{eq:dwdt}) and (\ref{eq:dsdt}) then
give
\begin{equation}
\label{eq:lalink2}
    \bar P-\bar E
    =
    \bar C
    =
    \overline{\mathrm{LQ}},
\end{equation}
where the overbar denotes a long-term average. Thus, net atmospheric
moisture convergence is balanced by terrestrial water export after
evapotranspiration.

Equation~(\ref{eq:lalink2}) connects the atmospheric and terrestrial
water balances. For a watershed with negligible long-term storage change
and small net groundwater exchange across its boundary, basin discharge
provides an observational approximation to terrestrial water export and
therefore to atmospheric moisture convergence.

\subsection{Role of Rainfall Frequency and Intensity}
\label{sec:mini}
To relate the long-term water balance to rainfall intermittency, we use a
stochastic ecohydrological model in which precipitation frequency and
event depth enter explicitly \cite{porporato2004,ecohydrology2022}.
This distinction is essential because different combinations of rainfall
frequency and intensity can produce the same mean precipitation while
leading to different partitioning between evapotranspiration and
hydrological export.

Rainfall intermittency is characterized by the frequency of rainfall
occurrence, $\lambda$, and the mean precipitation depth per event,
$\alpha$. The long-term mean precipitation rate is therefore
\begin{equation}
\label{eq:palla}
    \bar P=\lambda\alpha,
\end{equation}
with $[\bar P]=\mathrm{LT}^{-1}$,
$[\lambda]=\mathrm{T}^{-1}$, and $[\alpha]=\mathrm{L}$.
Specifically, rainfall arrivals are represented as a compound Poisson process with
rate $\lambda$, while event depths are exponentially distributed with
mean $\alpha$ \cite{porporato2004,ecohydrology2022}. Rainfall exceeding
the available soil-water storage contributes to runoff and deep drainage,
whereas evapotranspiration increases linearly with relative soil moisture
$s$ up to the potential rate $E_{\max}$. Within each climatological
state, $E_{\max}$ is represented by its long-term mean value; short-term
fluctuations in potential evaporation have a limited effect on the
long-term soil-water balance \cite{Daly2006}. The climatological value
of $E_{\max}$ may nevertheless vary between climatic states, as allowed
in the analysis below.

Under statistical steady state, the resulting evaporative fraction is
\cite{porporato2004,Daly2019,ecohydrology2022}
\begin{equation}
\label{eq:budyko}
    \frac{\bar E}{\bar P}
    =
    f(D_I,\gamma)
    =
    1
    -
    D_I
    \frac{
    \gamma^{\gamma/D_I-1}e^{-\gamma}
    }
    {
    \Gamma(\gamma/D_I)
    -
    \Gamma(\gamma/D_I,\gamma)
    },
\end{equation}
where $\Gamma(\cdot)$ and $\Gamma(\cdot,\cdot)$ are the complete and
upper incomplete gamma functions, respectively. The two dimensionless
indices are
\begin{align}
\label{eq:DI}
    D_I
    &=
    \frac{E_{\max}}{\bar P}
    =
    \frac{E_{\max}}{\alpha\lambda},
    \\
\label{eq:gamma}
    \gamma
    &=
    \frac{w_0}{\alpha}.
\end{align}

The dryness index $D_I$ measures atmospheric evaporative demand relative
to mean precipitation supply, whereas the storage index $\gamma$ measures
effective water-storage capacity relative to mean rainfall-event depth.
The latter is central to the present problem. Because
$\gamma=w_0/\alpha$, a change in rainfall intensity modifies
hydrological partitioning even when mean precipitation is unchanged.
By contrast, conventional formulations written solely as $f=f(D_I)$
retain frequency and intensity only through their product
$\bar P=\alpha\lambda$.

The evaporative fraction generally increases with both $D_I$ and
$\gamma$. Increasing $D_I$ moves the system toward stronger water
limitation, whereas increasing $\gamma$ allows a larger fraction of an
individual rainfall event to be retained in the terrestrial reservoir
rather than exported directly through runoff or deep drainage, increasing
the opportunity for subsequent evapotranspiration.

Daily precipitation records, by contrast, provide substantially broader spatial and temporal coverage than sub-daily event records and exhibit statistical features similar to those at the event scale \cite{Morbidelli2020, MartinezVillalobos2019}. At daily resolution, $\lambda$ is interpreted as wet-day frequency and $\alpha$ as mean wet-day precipitation depth, so that Eq.~(\ref{eq:palla}) remains exact for the corresponding daily statistics. We therefore use daily frequency and intensity as observational counterparts of the event-scale variables in the stochastic model.

\section{Theoretical Development of Temperature Sensitivities of Precipitation Frequency and Intensity}

In this section, we derive the relation between the temperature sensitivities of precipitation frequency and intensity and identify the hydrological and forcing terms
that govern it. We start with stochastic ecohydrological model with explicitly consideration of rainfall-intensity effects on hydrological export and then discuss possible outcomes without consideration of this rainfall-intensity dependence.

\subsection{Temperature Sensitivity with Explicit Rainfall-Intensity Dependence}
In the long-term statistical steady-state limit of
Eq.~(\ref{eq:lalink2}), substituting Eq.~(\ref{eq:budyko}) and using
Eq. \eqref{eq:palla}  gives
\begin{equation}
\label{eq:master}
    \bar C
    =
    \alpha\lambda\,[1-f(D_I,\gamma)] .
\end{equation}

We allow the climatological quantities
$\alpha$, $\lambda$, $\bar C$, $E_{\max}$, and $w_0$
to vary with climatological mean temperature $T$ and use a prime to
denote $d/dT$. From Eqs.~(\ref{eq:DI}) and~(\ref{eq:gamma}),
\begin{equation}
\label{eq:fracDI}
    \frac{D_I'}{D_I}
    =
    \frac{E_{\max}'}{E_{\max}}
    -
    \frac{\alpha'}{\alpha}
    -
    \frac{\lambda'}{\lambda},
\end{equation}
and
\begin{equation}
\label{eq:fracgamma}
    \frac{\gamma'}{\gamma}
    =
    \frac{w_0'}{w_0}
    -
    \frac{\alpha'}{\alpha}.
\end{equation}

Differentiating Eq.~(\ref{eq:master}) with respect to $T$ and collecting
terms gives
\begin{align}
\label{eq:responsecompact}
&
\left(1-f+D_I f_{D_I}\right)
\left(
\frac{\lambda'}{\lambda}
+
\frac{\alpha'}{\alpha}
\right)
+
\gamma f_\gamma\frac{\alpha'}{\alpha}
\nonumber\\
&\qquad =
\frac{\bar C'}{\bar P}
+
f_{D_I}\frac{E_{\max}'}{\bar P}
+
f_\gamma\frac{w_0'}{\alpha},
\end{align}
where $f_{D_I}=\partial f/\partial D_I$ and
$f_\gamma=\partial f/\partial\gamma$. Equation~(\ref{eq:responsecompact}) makes explicit the different roles
of rainfall frequency and intensity. The combination
$\lambda'/\lambda+\alpha'/\alpha$ is the fractional change in mean
precipitation, $\bar P=\alpha\lambda$, whereas rainfall intensity has
the additional contribution
$\gamma f_\gamma\,\alpha'/\alpha$ because changing $\alpha$ also changes
the storage index $\gamma=w_0/\alpha$. Thus, frequency and intensity are
hydrologically equivalent only when the evaporative fraction is
insensitive to $\gamma$.

Solving Eq.~(\ref{eq:responsecompact}) for the fractional change in
rainfall frequency yields
\begin{equation}
\label{eq:hc}
    \frac{\lambda'}{\lambda}
    =
    \underbrace{
    \left[
    -1-
    \frac{\gamma f_\gamma}
    {1-f+D_I f_{D_I}}
    \right]}_{\displaystyle \beta_h}
    \frac{\alpha'}{\alpha}
    +
    \underbrace{
    \frac{
    \bar C'/\bar P
    +
    f_{D_I}E_{\max}'/\bar P
    +
    f_\gamma w_0'/\alpha}
    {1-f+D_I f_{D_I}}
    }_{\displaystyle \beta_a}.
\end{equation}
Equation~(\ref{eq:hc}) is the central analytical result of this study. Details of the differentiation leading to
Eq.~(\ref{eq:responsecompact}) are given in Appendix \ref{app:derivation}.
The dimensionless coefficient $\beta_h$ is the \emph{hydrological sensitivity}. It depends only on the
hydroclimatic state through $D_I$ and $\gamma$. For the stochastic
water-balance model considered here, $f_\gamma>0$ over the physically
relevant parameter range, so that generally $\beta_h<-1$. The departure
from $-1$ therefore arises from the explicit dependence of hydrological
partitioning on rainfall event depth.
The additive term $\beta_a$
represents changes in the forcing of the coupled water balance through
atmospheric moisture convergence, potential evapotranspiration, and
effective water-storage capacity. It is therefore not purely atmospheric
by definition, although the observational analysis below shows that its
spatial variability is dominated by changes in moisture convergence.

\begin{figure}
  \centering
  \includegraphics[width=0.8\linewidth]{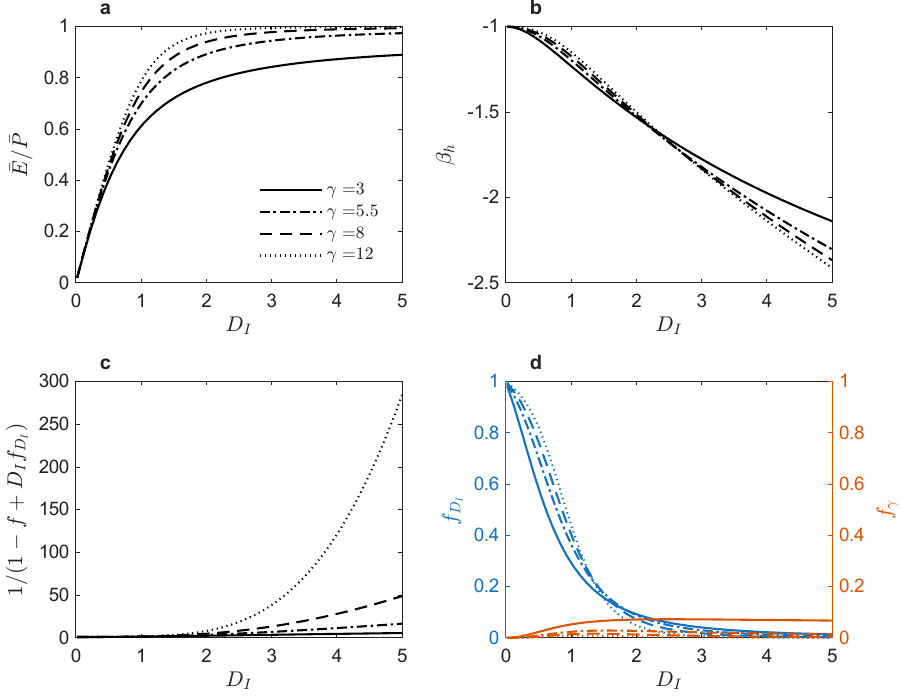}
  \caption{
  Hydrological partitioning and sensitivities predicted by the stochastic
  water-balance model. (a) Long-term evaporative fraction
  $\bar E/\bar P=f(D_I,\gamma)$; (b) hydrological sensitivity $\beta_h$;
  (c) amplification factor
  $[1-f+D_I f_{D_I}]^{-1}$ appearing in the forcing term; and
  (d) derivatives $f_{D_I}$ and $f_\gamma$, shown as functions of
  dryness index $D_I$ for different values of the storage index $\gamma$.
  }
  \label{fig:budhc}
\end{figure}

Figure~\ref{fig:budhc}b shows that $\beta_h$ generally becomes more
negative with increasing dryness. This behavior reflects the increasing
importance of rainfall event depth for hydrological export under
water-limited conditions. The dependence on storage is non-monotonic
with $D_I$, with the curves crossing near $D_I\simeq2.1$.

Equation~(\ref{eq:hc}) therefore gives a physical origin to the negative
frequency--intensity slope. For fixed $w_0$, increasing rainfall
intensity decreases $\gamma=w_0/\alpha$, reducing the evaporative
fraction and increasing the fraction of precipitation exported from the
terrestrial reservoir. A larger compensating decrease in rainfall
frequency is consequently required for a given forcing. The resulting
negative slope is a hydrological constraint; the forcing term determines
the displacement of the relation.

The limiting cases help clarify this behavior. In the humid limit
($D_I\rightarrow0$) and in the vanishing-storage limit
($\gamma\rightarrow0$), the hydrological sensitivity approaches
$\beta_h=-1$, so that frequency and intensity become hydrologically
equivalent at leading order. In contrast, in the hyperarid limit
($D_I\rightarrow\infty$ at fixed $\gamma$),
$\beta_h\rightarrow-1-\gamma$, while the forcing term is increasingly
amplified as the storage index grows. Thus, the departure from $-1$ is
most pronounced under dry conditions where finite storage makes event
depth important for hydrological partitioning. These and the
large-$\gamma$ limits are derived and discussed in Appendix \ref{app:limits}.

\subsection{Dryness-Index-Only Models and the Loss of Rainfall-Intensity Dependence}

The importance of explicit rainfall-intensity dependence becomes clear
for hydrological models in which the long-term evaporative fraction is
written only as
\begin{equation}
    f=f(D_I).
\end{equation}
Because the dryness index in Eq.~(\ref{eq:DI}) depends on precipitation
frequency and intensity only through their product
$\bar P=\alpha\lambda$, such formulations cannot distinguish between
rainfall regimes having the same mean precipitation but different
combinations of event frequency and event depth.

With no explicit storage-index dependence,
\begin{equation}
    f_\gamma=0,
\end{equation}
and the hydrological sensitivity becomes identically
\begin{equation}
\label{eq:limitbudyko}
    \beta_h=-1.
\end{equation}
A fractional change in rainfall intensity then has the same long-term
hydrological effect as the same fractional change in rainfall frequency.

This limitation arises because event depth affects runoff and deep
drainage through finite water-storage capacity. Two rainfall regimes
with the same $\bar P$ but different $\alpha$ and $\lambda$ may therefore
produce different long-term hydrological partitioning. A
dryness-index-only formulation contains no variable that measures event
depth relative to storage and cannot represent this effect.

In the stochastic model, that information is retained through
\begin{equation}
    \gamma=\frac{w_0}{\alpha},
\end{equation}
and the term $\gamma f_\gamma$ measures the additional hydrological
sensitivity associated with rainfall intensity. This term is precisely
what allows the frequency--intensity slope to depart from $-1$.

This case should be distinguished from holding $\gamma$ constant within
the full model $f(D_I,\gamma)$. If $\gamma'=0$, then
\begin{equation}
    \frac{w_0'}{w_0}
    =
    \frac{\alpha'}{\alpha},
\end{equation}
and the corresponding storage contribution cancels the
$\gamma f_\gamma$ contribution to the intensity response. The effective
frequency--intensity slope is again $-1$, even though
$f_\gamma\neq0$. Thus, $f_\gamma=0$ represents a structural absence of
rainfall-intensity dependence, whereas $\gamma'=0$ describes a
particular trajectory through a model that retains that dependence.
The distinction is derived explicitly in Appendix \ref{app:derivation}.

\section{Estimating the Forcing Components from Hydrometeorological Records}
\label{sec:forcingobs}

We estimate the components of the forcing term $\beta_a$ in
Eq.~(\ref{eq:hc}) using long-term hydrometeorological records. The
primary dataset is the Model Parameter Estimation Experiment (MOPEX),
which provides daily precipitation and streamflow records from 1948 to
2003 for 438 small- to medium-sized basins across the contiguous United
States with limited human regulation \cite{Duan2006}. Long-term
hydrological partitioning in these basins is well represented by the
stochastic water-balance model introduced in
Sec.~\ref{sec:wbframe} \cite{Daly2019,Daly2019wrr}. We complement MOPEX
with potential evapotranspiration from the Climatic Research Unit (CRU)
\cite{Harris2020} and global-mean temperature from GISTEMP v4
\cite{Hansen2010}. Together, these datasets allow the individual
contributions to $\beta_a$ to be estimated over a common period.

For convenience, we write
\begin{equation}
\label{eq:betaadecomp}
  \beta_a
  =
  \frac{
  \bar C'/\bar P
  +
  f_{D_I}E'_{\max}/\bar P
  +
  f_\gamma w'_0/\alpha
  }
  {1-f+D_I f_{D_I}}
  =
  \frac{a_t}{a_h}
  =
  \frac{a_1+a_2+a_3}{a_h},
\end{equation}
where
\begin{equation}
    a_h=1-f+D_I f_{D_I}
\end{equation}
is the hydrological denominator, and
\begin{equation}
    a_1=\frac{\bar C'}{\bar P},
    \qquad
    a_2=f_{D_I}\frac{E'_{\max}}{\bar P},
    \qquad
    a_3=f_\gamma\frac{w'_0}{\alpha}.
\end{equation}
The three terms in the numerator represent contributions from changes
in atmospheric moisture convergence, potential evapotranspiration, and
effective water-storage capacity, respectively.

The factor $1/a_h$ determines how strongly these forcing contributions
are expressed in $\beta_a$. As shown in Fig.~\ref{fig:budhc}c, the
amplification increases strongly toward dry conditions and also depends
on the storage index. For example, at $D_I=5$, $1/a_h$ ranges from about
5.6 for $\gamma=3$ to about 280 for $\gamma=12$. Thus, the same forcing
in the numerator may produce a much larger displacement of the
frequency--intensity relation in dry regimes with large storage index.

We first estimate the moisture-convergence contribution $a_1$. For each
MOPEX basin, streamflow is averaged over a 15-year moving window advanced
annually through the 1948--2003 record. Over these long intervals, and
neglecting net changes in terrestrial water storage and groundwater
exchange, basin discharge is used as an observational approximation to
atmospheric moisture convergence through Eq.~(\ref{eq:lalink2}). The
resulting window-averaged streamflow series is regressed against the
corresponding 15-year mean global temperature from GISTEMP. The
regression slope, divided by the long-term mean precipitation of the
basin, provides the estimate of $a_1$. Windows containing missing
records are excluded, and only basins with at least 20 valid window
averages are retained, leaving 384 basins for subsequent analysis. The main results are robust to the choice of averaging window, with similar findings for 10- and 20-yr windows.

Estimation of $a_2$ and $a_3$ requires the basin dryness and storage
indices. The long-term dryness index,
\begin{equation}
    D_I=\frac{E_{\max}}{\bar P},
\end{equation}
is calculated from CRU potential evapotranspiration and MOPEX
precipitation over 1948--2003. Long-term evapotranspiration is estimated
from the basin water balance as precipitation minus streamflow,
neglecting long-term storage change. The storage index $\gamma$ is then
obtained by inverting Eq.~(\ref{eq:budyko}) using the long-term
$\bar E/\bar P$ ratio and the estimated $D_I$.

The potential-evapotranspiration contribution $a_2$ is obtained by first
evaluating $f_{D_I}$ at the estimated $D_I$ and $\gamma$. CRU potential
evapotranspiration is averaged over the same 15-year moving windows and
regressed against global-mean temperature to estimate $E'_{\max}$.
Together with the long-term mean precipitation, this yields
\begin{equation}
    a_2=f_{D_I}\frac{E'_{\max}}{\bar P}.
\end{equation}

For the storage contribution $a_3$, we similarly evaluate $f_\gamma$
from the estimated hydrological state. Within each 15-year window, the
Budyko relation is inverted to obtain a window-specific effective storage
index $\gamma$, and
\begin{equation}
    w_0=\gamma\alpha
\end{equation}
is used to estimate the corresponding effective storage capacity.
Regression of these values against global-mean temperature provides
$w'_0$, from which
\begin{equation}
    a_3=f_\gamma\frac{w'_0}{\alpha}
\end{equation}
is obtained. Here $w_0$ is interpreted as an effective hydrological
storage parameter inferred from the long-term water balance rather than
as a direct measurement of physical soil-water capacity.

To quantify the relative importance of the three contributions, we
decompose the spatial variance of $a_t=a_1+a_2+a_3$ as
\begin{align}
\label{eq:vardecomp}
  \mathrm{Var}(a_t)
  ={}&
  \mathrm{Var}(a_1)
  +\mathrm{Var}(a_2)
  +\mathrm{Var}(a_3)
  \nonumber\\
  &+
  2\mathrm{Cov}(a_1,a_2)
  +2\mathrm{Cov}(a_1,a_3)
  +2\mathrm{Cov}(a_2,a_3).
\end{align}

\begin{figure}
  \centering
  \includegraphics[width=0.5\linewidth]{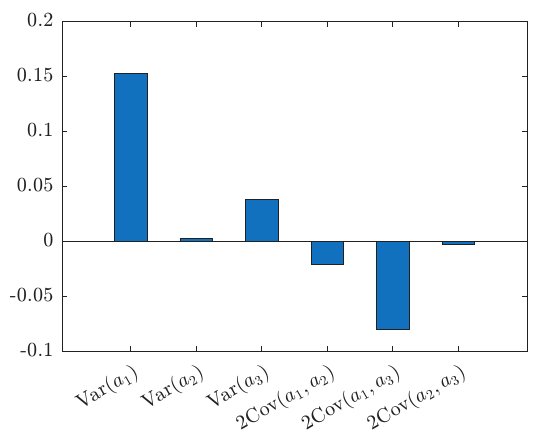}
  \caption{
  Variance decomposition of the forcing numerator
  $a_t=a_1+a_2+a_3$ across the 384 MOPEX basins.
  Negative covariance terms reduce the total variance.
  }
  \label{fig:var}
\end{figure}

The results for the 384 basins are shown in Fig.~\ref{fig:var}.
Variability in $a_1$ is by far the largest contribution, indicating that
changes in moisture convergence dominate the spatial variability of the
forcing numerator. The contribution from potential evapotranspiration is
much smaller, reflecting both the relatively weak variability of
$E'_{\max}$ and the small magnitude of $f_{D_I}$
(Fig.~\ref{fig:budhc}d). The effective-storage contribution is also
comparatively small because variations in $w'_0$ are multiplied by the
generally small sensitivity $f_\gamma$.

Among the covariance terms, $\mathrm{Cov}(a_1,a_3)$ provides the largest
secondary contribution and is negative, thereby reducing the total
variance. This behavior is consistent with increasing effective storage
reducing the efficiency of hydrological export and therefore partly
opposing changes in moisture convergence. Overall, the dominance of
$a_1$ indicates that variability in the forcing term is predominantly
associated with changes in atmospheric moisture convergence in the
MOPEX basins, motivating the interpretation of $\beta_a$ as an
atmospherically dominated forcing term in the analysis below.

\section{Temperature Sensitivities of Precipitation Frequency and Intensity Across Hydrological Regimes}
\label{sec:regimes}

We next examine how the temperature sensitivities of precipitation
frequency and intensity vary across hydrological regimes. According to
Eq.~(\ref{eq:hc}), their covariation has a slope determined by the
hydrological sensitivity $\beta_h$ and an intercept determined by the
forcing term $\beta_a$. To isolate the dependence of the slope on
hydrological state, we compare MOPEX basins in wet ($D_I<1$) and dry
($D_I>1$) regimes while restricting the forcing term to the common range
\begin{equation}
    0.1\; \mathrm{K}^{-1} <\beta_a<0.2\;\mathrm{K}^{-1}.
\end{equation}

\begin{figure}
  \centering
  \includegraphics[width=0.5\linewidth]{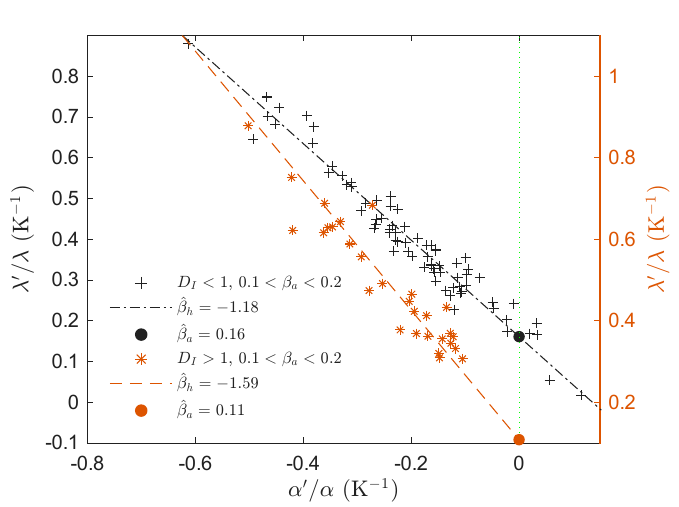}
  \caption{
  Temperature sensitivities of precipitation frequency and intensity
  across wet and dry hydrological regimes. Points show
  $\lambda'/\lambda$ and $\alpha'/\alpha$ for MOPEX basins with
  $0.1<\beta_a<0.2$, separated into wet ($D_I<1$) and dry
  ($D_I>1$) regimes. Lines denote the first principal components of the
  two distributions. Their slopes provide estimates of the hydrological
  sensitivity, $\hat{\beta}_h$, and are steeper in the dry regime, as
  predicted by Eq.~(\ref{eq:hc}).
  }
  \label{fig:anti2}
\end{figure}

The resulting temperature sensitivities,
$\lambda'/\lambda$ and $\alpha'/\alpha$, are shown in
Fig.~\ref{fig:anti2}. A pronounced negative covariation is evident in
both regimes, but the relation is substantially steeper under dry
conditions. The first principal component (PC1) has slope
$\hat{\beta}_h=-1.18$ for the wet basins and
$\hat{\beta}_h=-1.59$ for the dry basins.

These values are consistent with the hydrological sensitivity predicted
by Eq.~(\ref{eq:hc}). For representative values $D_I=0.95$ and
$\gamma=5.5$ in the wet regime, the theory gives
$\beta_h\simeq-1.18$, whereas for $D_I=2.21$ and $\gamma=5.5$ in the
dry regime it gives $\beta_h\simeq-1.59$
(Fig.~\ref{fig:budhc}b). The observed steepening of the
frequency--intensity covariation from wet to dry conditions is therefore
consistent with the dependence of $\beta_h$ on hydrological state
predicted by the stochastic water-balance model.

The corresponding PC1 intercepts are 0.16 and 0.11 for the wet and dry
regimes, respectively, both within the prescribed range of $\beta_a$.
Because the basins were selected within this range, the intercepts are
not an independent test of the theory. Rather, restricting $\beta_a$
reduces variation in the forcing term and allows the dependence of the
slope on hydrological state to be examined more directly. The remaining
scatter around the PC1 lines reflects variations in hydrological
properties, particularly the storage index $\gamma$, together with
residual variability in $\beta_a$ within the selected range.

We use principal-component analysis because uncertainties affect both
$\lambda'/\lambda$ and $\alpha'/\alpha$. Ordinary least-squares
regression treats the predictor variable as error-free and may therefore
bias the estimated slope when both variables contain uncertainty
\cite{Hutcheon2010}. More importantly, an unconditional regression
across all basins would combine variations in the hydrological
sensitivity $\beta_h$ with variations in the forcing term $\beta_a$.
In particular, increases in rainfall intensity may be accompanied by
increases in moisture convergence, producing a correlation between
$\alpha'/\alpha$ and $\beta_a$ that can be absorbed into the fitted
frequency--intensity slope. The corresponding correlation between
$\lambda'/\lambda$ and $\beta_a$ is much weaker. Reversing the
regression therefore does not eliminate the problem and remains subject
to uncertainty in both variables.

Conditioning on a restricted range of $\beta_a$ reduces this confounding
and provides a more direct comparison between the observed
frequency--intensity covariation and the hydrological sensitivity
predicted by the theory. The contrast between the wet and dry regimes
therefore supports the central result of the framework: the temperature
sensitivities of precipitation frequency and intensity are linked by the
long-term water balance, while the slope of their relationship depends
systematically on hydrological state.

\section{Conclusions}

We developed an analytical framework linking the temperature
sensitivities of precipitation frequency and intensity through the
coupled land--atmosphere water balance. By combining this balance with a
stochastic hydrological model that retains explicit dependence on both
rainfall frequency and event depth, we obtained
\begin{equation}
    \frac{\lambda'}{\lambda}
    =
    \beta_h\frac{\alpha'}{\alpha}
    +
    \beta_a .
\end{equation}
The slope $\beta_h$ is a hydrological sensitivity determined by the
dryness and storage indices, whereas $\beta_a$ represents changes in the
forcing of the coupled water balance through moisture convergence,
potential evapotranspiration, and effective water-storage capacity.

A central result is that rainfall frequency and intensity are not
hydrologically equivalent, even when their product
$\bar P=\alpha\lambda$ is unchanged. Changes in event depth also modify
the storage index $\gamma=w_0/\alpha$ and therefore the partitioning of
precipitation between evapotranspiration and hydrological export.
Dryness-index-only models cannot represent this effect and necessarily
give $\beta_h=-1$. In the stochastic formulation, the additional
dependence on $\gamma$ allows the slope to depart from $-1$ and generally
become more negative as conditions become drier.

Analysis of long-term MOPEX records supports this interpretation. The
variability of the forcing term is dominated by atmospheric moisture
convergence, whereas contributions from potential evapotranspiration and
effective storage changes are smaller. When basins are compared over a
restricted range of $\beta_a$, the observed frequency--intensity
covariation is steeper in dry than in wet regimes, consistent with the
predicted dependence of $\beta_h$ on hydrological state. Hydrological
partitioning therefore constrains the slope of the temperature response,
while atmospheric moisture supply provides the dominant contribution to
its displacement in the basins examined here.

More broadly, the framework complements statistical analyses of
land--atmosphere coupling by imposing a physical water-balance
constraint. Correlations among atmospheric and land-surface variables
can be difficult to interpret when atmospheric forcing dominates their
variability; here, the analytical structure 
identifies the distinct terms that must contribute to the observed
frequency--intensity response.

Several extensions are possible. Soil-moisture effects on boundary-layer
development, convective instability, and storm organization could be
introduced by allowing rainfall intensity to respond explicitly to land
surface state \cite{Yin2015,Cerasoli2021}. Nonlocal land--atmosphere
interactions could be incorporated by relating moisture convergence to
surface latent and sensible heat fluxes and the resulting mesoscale
circulation \cite{GarciaCarreras2011,Guillod2015,Giles2023}. Finally,
retaining finite changes in atmospheric and terrestrial water storage
would extend the framework from long-term climatic responses toward
seasonal and shorter timescales.

\appendix

\section{Derivation of the Frequency--Intensity Relation}
\label{app:derivation}
Differentiation of the long-term water balance of Eq. (\ref{eq:master}) with respect to climatological mean temperature gives,
after division by $\bar P=\alpha\lambda$,
\begin{equation}
\label{eq:appdiff}
\frac{\bar C'}{\bar P}
=
(1-f)
\left(
\frac{\lambda'}{\lambda}
+
\frac{\alpha'}{\alpha}
\right)
-
f_{D_I}D_I'
-
f_\gamma\gamma'.
\end{equation}
Using
\begin{equation}
D_I'
=
D_I
\left(
\frac{E_{\max}'}{E_{\max}}
-
\frac{\alpha'}{\alpha}
-
\frac{\lambda'}{\lambda}
\right),
\qquad
\gamma'
=
\gamma
\left(
\frac{w_0'}{w_0}
-
\frac{\alpha'}{\alpha}
\right),
\end{equation}
and collecting the frequency- and intensity-dependent terms yields
\begin{equation}
\label{eq:appcompact}
A_\lambda\frac{\lambda'}{\lambda}
+
A_\alpha\frac{\alpha'}{\alpha}
=
\frac{\bar C'}{\bar P}
+
f_{D_I}\frac{E_{\max}'}{\bar P}
+
f_\gamma\frac{w_0'}{\alpha},
\end{equation}
where
\begin{equation}
\label{eq:appresponses}
A_\lambda
=
1-f+D_I f_{D_I},
\qquad
A_\alpha
=
A_\lambda+\gamma f_\gamma .
\end{equation}
Therefore
\begin{equation}
\label{eq:apphc}
\frac{\lambda'}{\lambda}
=
-\frac{A_\alpha}{A_\lambda}
\frac{\alpha'}{\alpha}
+
\frac{
\bar C'/\bar P
+
f_{D_I}E_{\max}'/\bar P
+
f_\gamma w_0'/\alpha
}
{A_\lambda},
\end{equation}
which gives Eq.~(\ref{eq:hc}) with
$\beta_h=-A_\alpha/A_\lambda$.

\subsection{Hydrological response coefficients}

The two coefficients in Eq.~(\ref{eq:appresponses}) have a direct
interpretation. At fixed $E_{\max}$ and $w_0$,
\begin{equation}
\label{eq:partialresponses}
A_\lambda
=
\frac{1}{\bar P}
\frac{\partial\bar C}{\partial\ln\lambda},
\qquad
A_\alpha
=
\frac{1}{\bar P}
\frac{\partial\bar C}{\partial\ln\alpha}.
\end{equation}
Their difference,
\begin{equation}
\label{eq:response_difference}
A_\alpha-A_\lambda=\gamma f_\gamma,
\end{equation}
is the additional hydrological response associated with changing event
depth. Thus
\begin{equation}
\beta_h=-\frac{A_\alpha}{A_\lambda}
\end{equation}
is the ratio of the intensity and frequency responses of hydrological
export.

\subsection{Two cases giving a slope of -1}
\label{app:fixedgamma}

If hydrological partitioning has no explicit storage dependence,
$f=f(D_I)$ and hence $f_\gamma=0$. Equation~(\ref{eq:appresponses})
then gives $A_\alpha=A_\lambda$ and therefore
\begin{equation}
\beta_h=-1.
\end{equation}

The same effective slope arises along a trajectory of the full model
$f(D_I,\gamma)$ for which $\gamma$ remains constant. Since
$\gamma=w_0/\alpha$, the condition $\gamma'=0$ implies
\begin{equation}
\frac{w_0'}{w_0}
=
\frac{\alpha'}{\alpha}.
\end{equation}
The resulting storage contribution cancels the term
$\gamma f_\gamma$ in the intensity response of $A_\alpha$, leaving
frequency and intensity equivalent at fixed $\gamma$. This should not
be confused with $f_\gamma=0$: the former specifies a trajectory
through $(D_I,\gamma)$ space, whereas the latter removes storage
dependence from the hydrological model itself.

\section{Limiting Behavior of the Hydrological Sensitivity}
\label{app:limits}

Here we summarize the asymptotic limits of the hydrological sensitivity
$\beta_h$ used in the main text. For convenience in the calculations, let
\begin{equation}
a=\frac{\gamma}{D_I},
\qquad
g=1-f
=
D_I
\frac{\gamma^{a-1}e^{-\gamma}}
{\Gamma(a)-\Gamma(a,\gamma)} .
\end{equation}

\subsection{Humid and hyperarid limits}

For fixed $\gamma>0$, as $D_I\rightarrow0$ we have
$a\rightarrow\infty$. The lower incomplete gamma function is dominated
by its upper endpoint,
\begin{equation}
\Gamma(a)-\Gamma(a,\gamma)
\sim
\frac{\gamma^a e^{-\gamma}}{a-\gamma},
\end{equation}
which gives
\begin{equation}
g\sim1-D_I,
\qquad
f\sim D_I,
\qquad
f_{D_I}\rightarrow1,
\qquad
\gamma f_\gamma\rightarrow0 .
\end{equation}
Consequently,
\begin{equation}
\label{eq:DI0_beta}
\lim_{D_I\rightarrow0}\beta_h=-1.
\end{equation}

In the opposite limit, $D_I\rightarrow\infty$ at fixed $\gamma$,
$a\rightarrow0^+$. Using
\begin{equation}
\Gamma(a)-\Gamma(a,\gamma)
\sim \frac{\gamma^a}{a},
\qquad a\rightarrow0^+,
\end{equation}
we obtain
\begin{equation}
g\rightarrow e^{-\gamma},
\qquad
D_I f_{D_I}\rightarrow0,
\qquad
\gamma f_\gamma\rightarrow\gamma e^{-\gamma}.
\end{equation}
Thus
\begin{equation}
1-f+D_I f_{D_I}\rightarrow e^{-\gamma},
\end{equation}
and hence
\begin{equation}
\label{eq:DIinf_beta}
\lim_{D_I\rightarrow\infty}\beta_h=-1-\gamma
\end{equation}
The same result gives the corresponding amplification of the forcing
term,
\begin{equation}
\frac{1}{1-f+D_I f_{D_I}}
\rightarrow e^\gamma.
\end{equation}
In this hyperarid limit, event depth has a pronounced influence on the fraction of precipitation that contributes to hydrological export; the frequency--intensity trade-off therefore steepens as the storage index increases.

\subsection{Vanishing-storage limit}

For fixed $D_I>0$ and $\gamma\rightarrow0$, both
$a=\gamma/D_I$ and the upper limit of the lower incomplete gamma
function vanish. Writing
\begin{equation}
\Gamma(a)-\Gamma(a,\gamma)
=
\gamma^a
\int_0^1 u^{a-1}e^{-\gamma u},du
\end{equation}
gives, to leading order,
\begin{equation}
\Gamma(a)-\Gamma(a,\gamma)
\sim \frac{\gamma^a}{a}.
\end{equation}
It follows that
\begin{equation}
f=\gamma+O(\gamma^2),
\qquad
D_I f_{D_I}\rightarrow0,
\qquad
\gamma f_\gamma\rightarrow0,
\end{equation}
and therefore
\begin{equation}
\label{eq:gamma0beta}
\lim_{\gamma\rightarrow0}\beta_h=-1
\end{equation}

\subsection{Large-storage limit}

The limit $\gamma\rightarrow\infty$ depends on whether the system is
energy- or water-limited.

For $0<D_I<1$, the endpoint approximation gives
\begin{equation}
f=D_I+O(\gamma^{-1}),
\end{equation}
so that
\begin{equation}
f_{D_I}\rightarrow1,
\qquad
\gamma f_\gamma\rightarrow0,
\qquad
1-f+D_I f_{D_I}\rightarrow1.
\end{equation}
Hence
\begin{equation}
\lim_{\gamma\rightarrow\infty}\beta_h=-1,
\qquad 0<D_I<1.
\end{equation}

For $D_I>1$, the complementary asymptotic form gives
\begin{equation}
\label{eq:glargegamma}
g
\sim
\frac{
D_I^a e^{-a(D_I-1)}
}
{\sqrt{2\pi a}},
\qquad
a=\frac{\gamma}{D_I}.
\end{equation}
Logarithmic differentiation then yields
\begin{equation}
\gamma f_\gamma
\sim
g\left[
\frac{\gamma(D_I-1-\ln D_I)}{D_I}
+\frac12
\right],\quad
1-f+D_I f_{D_I}
\sim
g\left[
\frac{\gamma\ln D_I}{D_I}
+\frac12
\right].
\end{equation}
Their ratio therefore approaches
\begin{equation}
\frac{\gamma f_\gamma}
{1-f+D_I f_{D_I}}
\rightarrow
\frac{D_I-1-\ln D_I}{\ln D_I},
\end{equation}
and the large-storage limit is
\begin{equation}
\label{eq:gammainf_combined}
\lim_{\gamma\rightarrow\infty}\beta_h
=
\begin{cases}
-1, & 0<D_I \leq1, \\
-1-\dfrac{D_I-1-\ln D_I}{\ln D_I},
& D_I>1.
\end{cases}
\end{equation}
The two branches join continuously at $D_I=1$.

\bibliographystyle{unsrt}  
\bibliography{references}

\end{document}